\documentclass[twocolumn,showpacs,preprintnumbers,amsmath,amssymb]{revtex4}
\usepackage{graphicx}
\usepackage{dcolumn}
\usepackage{bm}

\begin{document}

\title{Effect of cavity photons on the generation of multi-particle
entanglement}
\author{Che-Ming Li$^1$}
\author{Yueh-Nan Chen$^{1}$}
\author{C.-W. Luo$^{1}$}
\author{Jin-Yuan Hsieh$^2$}
\author{Der-San Chuu$^{1}
\thanks{
Corresponding author email address: dschuu@mail.nctu.edu.tw;
Fax:886-3-5725230; Tel:886-3-5712121-56105.}$}
\altaffiliation[Corresponding author: ]{Fax:886-3-5725230, Tel:886-3-5712121-56105}
\email{dschuu@mail.nctu.edu.tw}
\affiliation{$^1$Institute and Department of Electrophysics, National Chiao Tung
University, Hsinchu 30050, Taiwan.}
\affiliation{$^2$Department of Mechanical Engineering, Ming Hsin University of Science
and Technology, Hsinchu 30401,Taiwan.}
\date{\today }

\begin{abstract}
A study on the cause of the multi-particle entanglement is presented in this
work. We investigate how dot-like single quantum well excitons, which are
independently coupled through a single microcavity mode, evolve into
maximally entangled state as a series of conditional measurements are taken
on the cavity field state. We first show how cavity photon affects the
entanglement purity of the double-exciton Bell state and the triple-exciton
W state. Generalization to multi-excitons W states is then derived
analytically. Our results pave the way for studying the crucial cause of
multi-particle collective effect.
\end{abstract}

\pacs{03.67.Mn,03.65.Xp,03.67.Lx}
\maketitle

The regulation methods of quantum information processing\cite{1} rely on
sharing maximally entangled pairs between distant parties. As it is well
known, the entangled pairs may become undesired mixed states due to
inevitable interactions with environments\cite{2}. For this reason, great
attentions have been focused on the agreement of entanglement purification%
\cite{3}, schemes of entanglement distillation\cite{4}, and the decoherence
mechanisms of quantum bits (qubits) in a reservoir\cite{5}.

The environment may play an active role on the formation of the nonlocal
effect under well considerations. Many investigations\cite{6} have been
devoted to the considerations of the reservoir-induced entanglement between
two remote qubits. By manipulating a third system which interacts with two
remote qubits\cite{7,8,9,10}, many schemes have been proposed to enhance the
entanglement fidelity. However, more general situations and considerations
are still lacking in a entanglement generation process, especially the
multi-particle entanglement generation. This issue is crucial for both the
real application of quantum communication and the study on the mechanism of
multi-particle entanglement.

In this paper, we study the mechanism of multi-particle entanglement
generation. The feasible physical system undertaken is dot-like single
quantum well exciton coupled through a single microcavity mode, which is
proposed and depicted in Fig.1. The whole procedure in the experimental
realization can be performed by optical initialization, manipulation, and
read-out of exciton state. In it, the qubit is coded in the presence of an
exciton in a quantum well (QW), namely the exciton state in $i$-th QW, $%
\left| e,h\right\rangle _{i}$, is considered as the logical state $\left|
1\right\rangle _{i}$, and the vacuum state, $\left| 0,0\right\rangle _{i}$,
which represents the state with no electron and hole, is coded as the
logical state $\left| 0\right\rangle _{i}$. To analyze the dynamics of the
multi-exciton entanglement, a series of conditional measurement can be taken
on the cavity field state by means of the electro-optic effect. First, we
demonstrate how double-exciton Bell state and the triple-exciton W state can
be generated under the effect of conditional measurements. Then we discuss
the cause of multi-exciton W state, and propose a general formulation of
entanglement generation further. Several essentials of effects of the cavity
photon will also be presented.

\begin{figure}[t]
\vspace*{0.8cm}\includegraphics[width=3.4cm,angle=-90]{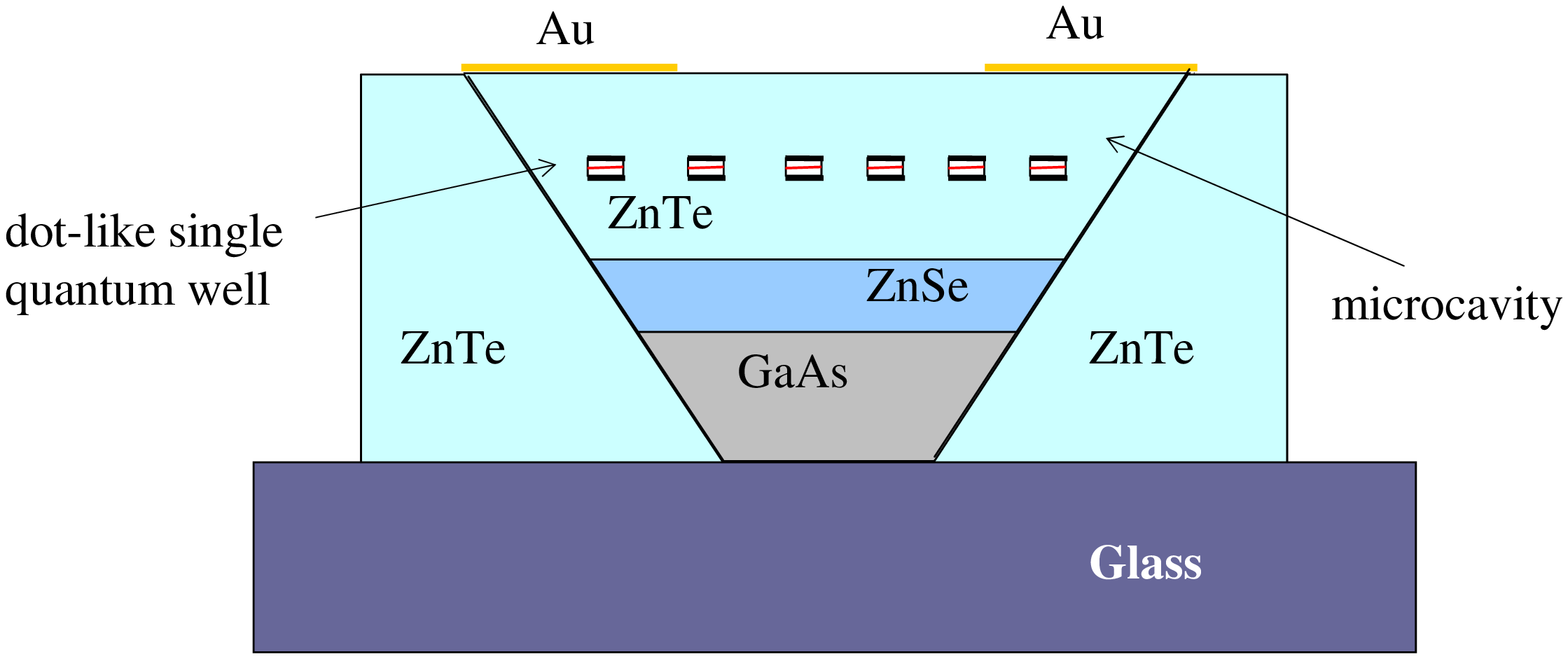}
\caption{The quantum devices with six dot-like quantum wells
inlaid in a microcavity which is constructed by a ZnTe medium and
two Au mirrors. This device can be prepared by the MBE, the e-beam
lithography, and the conventional semiconductor processing.}
\label{fig:epsart}
\end{figure}

In the QW-cavity system, we assume that the lateral size of the QWs are
sufficiently larger than the Bohr radius of excitons but smaller than the
wavelength of the photon fields. The dipole-dipole interactions and other
nonlinear interactions therefore can be neglected. The cavity mode is
assumed to be resonant with the excitons and equally interact with each QW.
Under the rotating wave approximation, the unitary time evolution of the $n$%
-QWs and cavity field is then governed by the interaction picture
Hamiltonian
\begin{equation}
H_{n(I)}=\sum_{m=1}^{n}\hbar \gamma (a\sigma _{m}^{+}+a^{+}\sigma _{m}^{-}),
\end{equation}
where $\gamma $ is the coupling constant, $a^{+}$ ($a$) is the
creation (annihilation) operator of the cavity field, and $\sigma
_{m}^{+}$ ($\sigma _{m}$) represents the creation (annihilation)
operator of the excitons in the $m$th QW.

For $n=2$ case and let the whole system be in the $m$-quanta
state, there would be an eigenstate of the Hamiltonian associated
with the photon trap, namely, $\left| \phi \right\rangle
=\frac{1}{\sqrt{2}}(\left| 1\right\rangle _{1}\left|
0\right\rangle _{2}-\left| 0\right\rangle _{1}\left|
1\right\rangle _{2})\otimes \left| m-1\right\rangle _{c},$ where
$\left| m-1\right\rangle _{c}$ refers to the cavity field state
with $m-1$ quanta. Once the system is in this state, the whole
system does not decay at all. Accordingly, keeping the cavity mode
in state $\left| m-1\right\rangle _{c}$ paves the way to generate
the entangled excitons, $\left| \psi \right\rangle
=\frac{1}{\sqrt{2}}(\left| 1\right\rangle _{1}\left|
0\right\rangle _{2}-\left| 0\right\rangle _{1}\left|
1\right\rangle _{2})$. For entanglement generation, the double
dot-like single QWs and cavity mode is prepared in the initial
state, $\left| \psi _{0}\right\rangle =\left| 1\right\rangle
_{1}\left| 0\right\rangle _{2}\left| 0\right\rangle _{c}$. For the
sake of generality and purpose of distillation, the state of QW
can be any mixed state, $\rho _{\psi }$, except the vacuum state.
Next, a pulse with $m$ photons is injected into the microcavity.
For the feasibility and the modest technology requirements, we
consider here the injection of single photon, i.e. $m=1$. The
total number of quantum count of the system is two. As the
single-photon has been injected into the cavity, the total system
will evolve with time, and the evolution operator $U(t)$ can be
easily derived form Eq. (1). If the system evolves without
interruption, it will go into a QW1-QW2-cavity field entangled
state. If we take a measurement on the cavity field state at some
instant, the number of the photon count of the detector may be
one, two, or zero. Since the single-photon state $\left|
1\right\rangle _{c}$ involves the photon-trapping phenomenon, we
can infer that the double-QW will evolve into a maximal entangled
state if the cavity mode stays in state $\left| 1\right\rangle
_{c}$ via the quantum jump approach\cite{11}. After measuring the
cavity field state, injecting a subsequent photon is necessary for
the sake of keeping the photon in its state. We then let the whole
system evolve for another period of time $\tau $ . Again, we
proceed to measure the cavity photon to make sure whether it is
one or not. If the cavity photon remains in single-photon, the
repetition continues; if not, the whole procedure should be
started over.

Therefore, after several times of successful repetitions, the state of
double-exciton progresses into the state, $\rho _{\psi
_{N}}=(_{c}\left\langle 1\right| U(\tau )\left| 1\right\rangle _{c})^{N}\rho
_{\psi _{i}}(_{c}\left\langle 1\right| U(-\tau )\left| 1\right\rangle
_{c})^{N}/P_{N,n=2}$, where $P_{N,n=2}=\text{Tr}[(_{c}\left\langle 1\right|
U(\tau )\left| 1\right\rangle _{c})^{N}\rho _{\psi _{i}}(_{c}\left\langle
1\right| U(-\tau )\left| 1\right\rangle _{c})^{N}]$, is the probability of
success for measuring a single-photon after $N$ times of repetitions. The
superoperator, $_{c}\left\langle 1\right| U(\tau )\left| 1\right\rangle _{c}$%
, reveals the significant fact that it will evolve to the
projection operator, $\left| \psi \right\rangle \left\langle \psi
\right| $, as the successful repetitions increases. This result
comes from the fact that the superoperator, $_{c}\left\langle
1\right| U(\tau )\left| 1\right\rangle _{c}$ , has only one
eigenvalue whose absolute value equals to one\cite{7,9} and the
corresponding eigenvector is just the photon-trapping state. Thus
the double-exciton state will become a maximal entangled state
after sufficient large repetitions. The probability of success and
fidelity can be evaluated as: $P_{N,2}=\frac{1}{2}(1+\cos
(\sqrt{6}\gamma \tau )^{2N})$ and $ F_{N,2}=\left\langle \psi
\right| \rho _{\psi _{N}}\left| \psi \right\rangle =\frac{1}{2\cos
(\sqrt{6}\gamma \tau )^{2N}}$. If $\gamma \tau $ is set to be
$(2m-1)\pi /(4\sqrt{6})$, $m=1,2,...,$ the fidelity of the
double-exciton state will approach to one and the probability of
success will be $1/2$ for large $ N$ . If the state of the cavity
field is kept in $m-$photon state, rather than the state with
single photon, the above results can be generalized to $
P_{N,2}=\frac{1}{2}(1+\cos (\sqrt{2(2m+1)}\gamma \tau )^{2N})$,
and $F_{N,2}= \frac{1}{2\cos (\sqrt{2(2m+1)}\gamma \tau )^{2N}}$.
It reveals that the number of measured cavity photon and the
evolution period play important roles in the trade-off between
$P_{N,2}$ and $F_{N,2}$. We can choose a set of $(m,\gamma \tau )$
such that the fidelity progresses to one at the least repetitions,
however, in the same time it causes the probability to reduce to a
minimum. On the other hand, one can also find a suitability such
that the probability goes to one. In this case, the system will
not evolve with time and is similar to the Zeno paradox with
\emph{finite} duration between two measurements.

We may directly follow the scheme based on continues measurements to achieve
the three-particle entanglement generation. However, we have observed that
the symmetry of the three-particle Hamiltonian is quite different from the
two-particle one, and one may hardly expect what kinds of entangled states
can be generated via the quantum jump method and even query whether the
quantum jump approach can pave the way for multi-particle entanglement
generation. In what follows we will first show that three-particle entangled
state indeed can be produced via conditional measurements; moreover the
W-type maximally entanglement can rise in multi-QWs system. The
superoperator that governs the progress of the three dot-like single QWs can
be worked out:
\begin{eqnarray}
_{c}\left\langle 1\right| e^{iH_{3(I)}\tau }\left| 1\right\rangle
_{c}&=&g\left| g\right\rangle \left\langle g\right| +W_{1}\left|
W_{1}\right\rangle \left\langle W_{1}\right| +T_{1}\left| T_{1}\right\rangle
\left\langle T_{1}\right|  \nonumber \\
&+& T_{2}\left| T_{2}\right\rangle \left\langle T_{2}\right|+ e\left|
e\right\rangle \left\langle e\right| +W_{2}\left| W_{2}\right\rangle
\left\langle W_{2}\right|  \nonumber \\
&+&T_{3}\left| T_{3}\right\rangle \left\langle T_{3}\right| +T_{4}\left|
T_{4}\right\rangle \left\langle T_{4}\right|,
\end{eqnarray}
where $g$, $e$, $W_{1}$, $W_{2}$, $T_{1}$, $T_{2}$, $T_{3}$ and $T_{4} $ are
functions of $\tau $ corresponding to the orthonormal eigenvectors
\begin{eqnarray}
&&\left| g\right\rangle =\left|000\right\rangle, \left| W_{1}\right\rangle=%
\frac{1}{\sqrt{3}}(\left|
100\right\rangle+\left|010\right\rangle+\left|001\right\rangle,  \nonumber \\
&&\left| T_{1}\right\rangle =\frac{1}{\sqrt{2}}(\left|
100\right\rangle-\left|001\right\rangle,  \nonumber \\
&&\left| T_{2}\right\rangle =\frac{1}{\sqrt{6}}(\left|
100\right\rangle-2\left|101\right\rangle+\left|001\right\rangle,  \nonumber
\\
&&\left| e\right\rangle=\left|111\right\rangle, \left| W_{2}\right\rangle =%
\frac{1}{\sqrt{3}}(\left|011\right\rangle
+\left|101\right\rangle+\left|110\right\rangle,  \nonumber \\
&&\left| T_{3}\right\rangle =\frac{1}{\sqrt{2}}(\left|
011\right\rangle-\left|110\right\rangle, \text{and }  \nonumber \\
&&\left| T_{4}\right\rangle =\frac{1}{\sqrt{6}}(\left|
011\right\rangle-2\left|010\right\rangle+\left|110\right\rangle).
\end{eqnarray}
Here $T_{1}=T_{2}$ and $T_{3}=T_{4}$ are two two-fold degenerate eigenvalues
of the superoperator $_{c}\left\langle 1\right| e^{iH_{3(I)}\tau }\left|
1\right\rangle _{c}$.

If the initial state is set equal to $\left| \psi _{0}\right\rangle
=\left|100\right\rangle$, the probability of success for finding the exciton
state, $\left| W_{1}\right\rangle$, can then be worked out analytically:
\begin{equation}
P_{N,3}=\frac{1}{3}(\cos (\sqrt{10}\gamma \tau )^{2N}+2\cos (\gamma \tau
)^{2N}),
\end{equation}
and the corresponding fidelity of the three-exciton state is
\begin{eqnarray}
F_{N,3} &=&\left\langle W_{1}\right| \rho _{\psi _{N}}\left|
W_{1}\right\rangle  \nonumber \\
&=&\frac{\cos (\sqrt{10}\gamma \tau )^{2N}}{\cos (\sqrt{10}\gamma \tau
)^{2N}+2\cos (\gamma \tau )^{2N}}.
\end{eqnarray}
Here, we can set $\gamma \tau $ to be $n\pi /\sqrt{10}$, $n=1,2,...$, then
the fidelity of the three-exciton state approaches to unit as $N$ increases;
meanwhile the probability of success is $1/3$.

\begin{figure}[t]
\vspace*{2.8cm} \includegraphics[width=2.7cm,angle=-90]{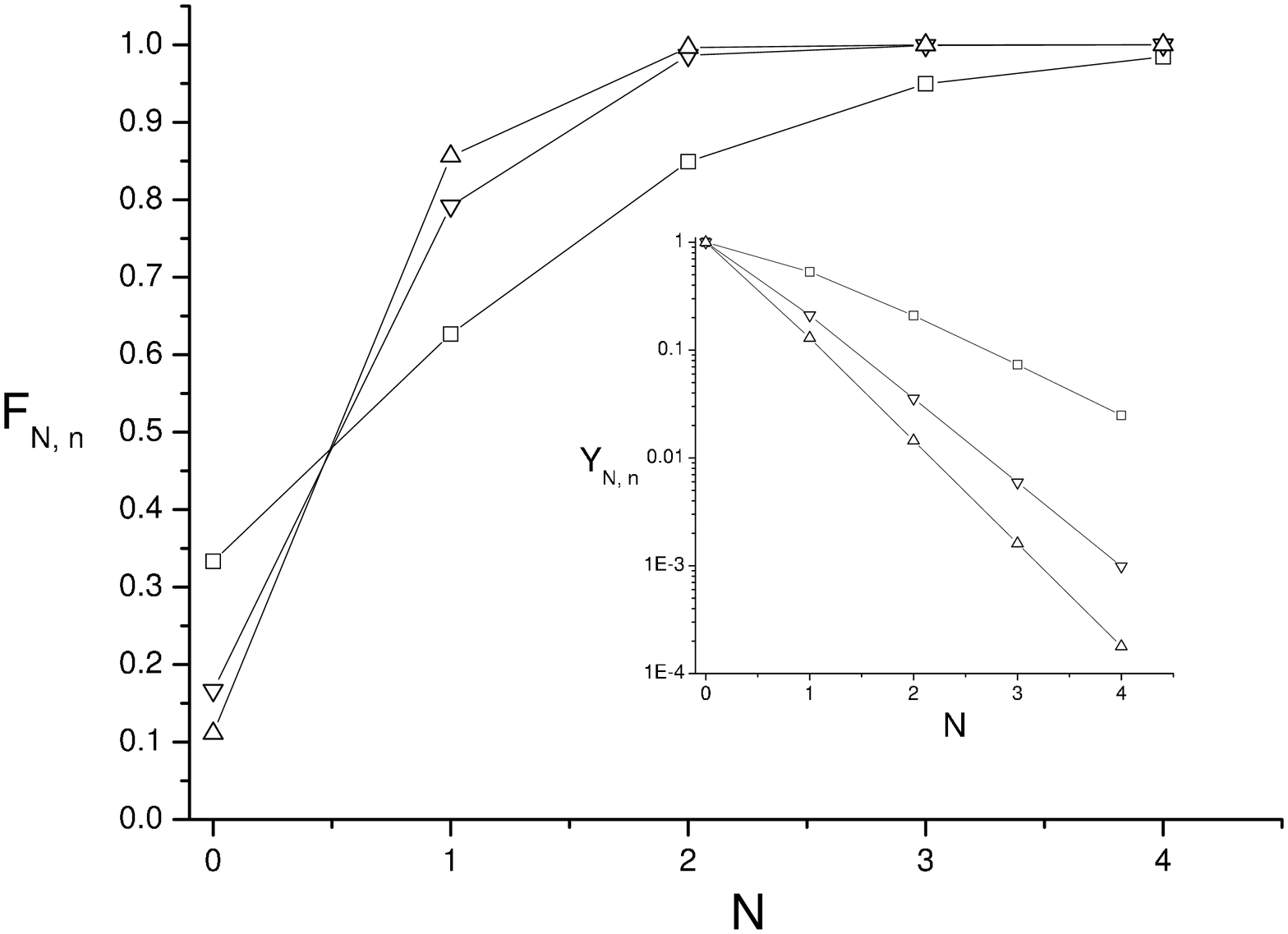}
\vspace*{1cm}
\caption{The variations of fidelity $F_{N,n}$ and the purification yield $%
Y_{N,n}$(in the inserted diagram) for cases $n=3(\Box), 6(\protect\nabla),$
and $9(\triangle)$, in which the evolution time of each case, $\protect\tau%
_{3}=\protect\pi/(\protect\sqrt{10}\protect\gamma)$, $\protect\tau_{6}=%
\protect\pi/(\protect\sqrt{22}\protect\gamma)$, and $\protect\tau_{9}=%
\protect\pi/(4\protect\sqrt{2}\protect\gamma)$ has been set.}
\label{fig:epsart}
\end{figure}

If the notion of Plenio et al.\cite{7} is generalized to the case of three
particles, in which a leak cavity is used for continuously measuring the
vacuum cavity field state $\left|0\right\rangle _{c}$, one can find that the
eigenvalue with unit norm of the superoperator, $_{c}\left\langle 1\right|
e^{iH_{3(I)}\tau }\left| 1\right\rangle _{c}$, is three-fold degenerate and
the set of corresponding eigenvectors is $\{\left|000\right\rangle,\left|
\psi_{ij} \right\rangle,\left| \psi_{kl} \right\rangle\}$, where $\left|
\psi_{ij(kl)} \right\rangle$ is the Bell-type state in which one photon is
trapped and shared between $i(k)$-QW and $j(l)$-QW, $i\neq j$, $k\neq l$ and
$i,j,k,l=1,2,3.$ Thus the collective motion cannot be induced via monitoring
the cavity vacuum.

Now we investigate the general case of multi-QWs entanglement generation
further. It is provable that the $n$-particle one-photon-trapping W state,
\begin{equation}
\left| \psi _{\text{W1}}\right\rangle =\frac{1}{\sqrt{n}}(\left|
100...0\right\rangle +\left| 010...0\right\rangle +...\left|
0...01\right\rangle ),
\end{equation}
and the $n$-particle $(n-1)$-photon-trapping W state,
\begin{equation}
\left| \psi _{\text{W2}}\right\rangle =\frac{1}{\sqrt{n}}(\left|
011...1\right\rangle +\left| 101...1\right\rangle +...\left|
1...10\right\rangle ),
\end{equation}
are eigenstates of $_{c}\left\langle 1\right| e^{iH_{n(I)}\tau }\left|
1\right\rangle _{c}$ by means of the symmetric properties of the system
Hamiltonian, operator algebra, and the Taylor expansions of the
superoperator $_{c}\left\langle 1\right| e^{iH_{n(I)}\tau }\left|
1\right\rangle _{c}$, where $n=3,4,...$. However, there exists the only
state, $\left| \psi _{\text{W1} }\right\rangle $, can be generated through
conditional measurements, and the selectivity is judged by the criterion of
unit norm of which eigenvalues. The eigenvalue of the state $\left| \psi _{%
\text{W1}}\right\rangle $ can be derived, namely, $\cos (\sqrt{4n-2}\gamma
\tau )$, and its absolute value will be one only at some nodes, viz $\sqrt{%
4n-2}\gamma \tau =m\pi $, where $m $ is a positive integer. For the case of $%
n=6$, we prepare the initial state of six-QWs to be $\left|
100000\right\rangle $, following the standard procedure of entanglement
generation mentioned above and measuring the cavity mode under the condition
$\gamma \tau =\pi /\sqrt{22}$, then the six-QWs will evolve to state $\left|
\psi _{\text{W1}}\right\rangle $ with the probability
\begin{equation}
P_{N,6}=\frac{1}{6}(\cos (\sqrt{22}\gamma \tau )^{2N}+5\cos (2\gamma \tau
)^{2N}),
\end{equation}
meanwhile, with the generation fidelity
\begin{equation}
F_{N,6}=\frac{\cos (\sqrt{22}\gamma \tau )^{2N}}{\cos (\sqrt{22}\gamma \tau
)^{2N}+5\cos (2\gamma \tau )^{2N}}.
\end{equation}%
In which, the time period between two successive measurements can be shorter
than the one in the three-exciton case. Besides, the values of $P_{N,6}$ and
$F_{N,6}$ approach to steady in fewer steps, but the purification yield will
be lower. From the trend of a decrease in success probability, the
limitation and efficiency will be an issue when one devises the quantum
strategy that can induce a collective motion in the multi-particle system.
The generalization of multi-QWs entanglement formulation has been concluded
as follows,
\begin{widetext}
\begin{eqnarray}
&&P_{N,n}=\frac{1}{n}(\cos (\sqrt{4n-2}\gamma \tau
)^{2N}+(n-1)\cos (\sqrt{n-2}\gamma \tau )^{2N}),
\text{and} \\
&&F_{N,n}=\frac{\cos (\sqrt{4n-2}\gamma \tau )^{2N}}{\cos
(\sqrt{4n-2}\gamma \tau )^{2N}+(n-1)\cos (\sqrt{n-2}\gamma \tau
)^{2N}},
\end{eqnarray}
\end{widetext}where $n=3,4,...$. Fig. 2 shows the variations of the
probability $P_{N,n}$ and the purification yield $Y_{N,n}$, defined by $%
Y_{N,n}=\prod_{i=0}^{N}P_{i,n}$, for the cases of $n=3,6,$ and $9$. It
implies that if there is a large dot-like single QW embedded inside the
single mode cavity, the collective motion will be hard to exist in the
system via conditionally measuring one single photon of the cavity mode.
Although as the particle number grows the probability decreases, even at one
step the whole particles can be entangled in the W-type state.

Several essentials of multi-QWs entanglement and the collective effect
should be expatiated here. First, the W-type entangled states are not the
eigenstates of the system, so that the explanation of the progression of
entangled QWs based on the quantum jump approach will be quite different
from the double-QWs case. Not merely the quantum jump method does elaborate
the entanglement formation in decoherence-free subspace, but it specifies
the general formulation of multi-particle entanglement. The symmetry of
double-QWs-cavity system causes QWs to evolve into a singlet state, hence
these states will maintain the coherence through sharing one single photon
between them. However, this relation will be broken as the cavity field mode
mediates interaction between more than two QWs, consequently another type of
entanglement exists in the system, namely the W state. The collective
effect, which results in the formation of W states, does not always arise in
a multi-particle system. If the conditional result of the measurement
depends upon the vacuum of the cavity field and the purification acts like
the two-particle scheme of Plenino et al.\cite{7}, the QWs will get into a
dilemma in which each QW can cooperate with any other QW in the cavity to
trap one photon in the same instant due to the zero-quanta cavity mode, that
is the superoperator $_{c}\left\langle 0\right| e^{iH_{n(I)}\tau }\left|
0\right\rangle _{c}$ has a $(n-1)$-fold degenerate eigenvalue, $1$, and the
pairs of photon-trapping states constitute the corresponding set of
orhtonormal eigenvectors. Since the degeneracy inevitably appears in the
system under frequently measurements on the vacuum of the cavity field, the
notion of Plenino et al. therefore may not be extended to multi-particle
entanglement generation. Thus, keeping the cavity mode to retain one photon
such that one may block the formation of pair-wise photon trap is the cause
of multi-QWs collective effect. Secondly, when the procedure of entanglement
generation ends up with a $n$-particle W state and zero cavity photon, then
the QWs and cavity field state will evolve under the dynamics, $\cos(\sqrt{n}%
\gamma t)\left| \psi _{\text{W1}}\right\rangle\left| 0\right\rangle
_{c}-i\sin(\sqrt{n}\gamma t)\left|\mathbf{0}\right\rangle\left|1\right%
\rangle _{c}$, where $\left|\mathbf{0}\right\rangle\equiv\left|0\right%
\rangle^{\otimes n}$. If the total number of quanta is two, the evolution of
the state, $\left| \psi _{\text{W1}}\right\rangle\left|1\right\rangle_{c}$,
becomes, $\cos(\sqrt{3n-2}\gamma t)\left| \psi _{\text{W1}%
}\right\rangle\left|1\right\rangle _{c}-i\sin(\sqrt{3n-2}\gamma t)(\sqrt{%
\frac{n}{3n-2}}\left|\mathbf{0}\right\rangle\left|2\right\rangle _{c}+\sqrt{%
\frac{2n-2}{3n-2}}\left|\phi\right\rangle\left|0\right\rangle _{c})$, where $%
\left|\phi\right\rangle\equiv\frac{1}{\sqrt{n(n-1)}}(\left|110...0\right%
\rangle+\left|1010...0\right\rangle
+...+\left|0...011\right\rangle)$. From which, unlike the
decoherence-free state in $n=2$ case, the generalized W states
suffer from Rabi oscillations with the frequencies proportional to
the square root of the QW numbers. Finally, the photon number of
the cavity field state plays a crucial role of the quantum
strategies, based on conditional measurements, that can perform
exotic quantum computation. One may accomplish the initialization
of the qubits by conditionally monitoring two-photon state of the
cavity mode. Since the vacuum state associates with the
eigenvector of the superoperator $_{c}\left\langle 2\right|
e^{iH_{n(I)}\tau }\left| 2\right\rangle _{c}$ with eigenvalue
$\frac{n(\cos (\sqrt{4n-2}\gamma \tau)+1)-1}{2n-1}$, under a well
control of evolution time $\tau $ the qubits can then be
initialized.

To summarize, a general formulation of multi-QWs entanglement generation
based on conditioned measurements has been proposed in this work. We
investigate the entanglement generation of dot-like single quantum well
excitons coupled through a single microcavity mode. We first consider how
the cavity photon affects the purity of the exciton state and the
purification efficiency in two- and three-particle protocol. The mechanism
of multi-QWs entanglement generation and the cause of collective effect then
have been investigated. Thus, we formulate the dynamics of W-type entangled
state by the Eqs. (10) and (11). Finally, some essentials of multi-QWs
entanglement and the collective effect, including the block of photon trap
and the advanced application to quantum computation, have also be studied.

This work is supported partially by the National Science Council, Taiwan
under the grant numbers NSC 92-2120-M-009-010 and NSC 93-2112-M-009-037.

\end{document}